\documentclass[11pt]{article}[12pt]
\usepackage[utf8]{inputenc}
\usepackage[english]{babel}
\usepackage{amsmath}
\usepackage{amssymb}
\usepackage{amsthm}
\usepackage{graphicx}
\usepackage{epstopdf}
\usepackage[affil-it]{authblk}
\usepackage{upgreek}

\usepackage{hyperref}
\hypersetup{
    colorlinks=true,
    linkcolor=red,
    citecolor=blue,
    urlcolor=blue
    }

\usepackage[a4paper, total={6in, 9.6in}]{geometry}

\def\R{\mathbb{R}}

\newcommand{\pd}[2]{\dfrac{\partial #1}{\partial #2}}

\newcommand{\der}[2]{\dfrac{\mathrm{d} #1}{\mathrm{d} #2}}
\newcommand{\bra}[1]{\langle #1 |}
\newcommand{\ket}[1]{| #1 \rangle}

\def\I{\mathbb{I}}      %identity

\begin{document}

\title{\textbf{Quantum Gate Generation in Two-Level Open Quantum Systems by Coherent and Incoherent Photons Found with Gradient Search}}

\date{}

\author[1,2,*]{Vadim N. Petruhanov}
\author[1,2,**]{Alexander N. Pechen}

\affil[1]{\it \normalsize \href{http://mi-ras.ru/eng/dep51
}{Department of Mathematical Methods for Quantum Technologies},\par
Steklov Mathematical Institute of Russian Academy of Sciences,\par
8~Gubkina str., Moscow, 119991, Russia, }
\affil[2]{\it Quantum Engineering Research and Education Center,\par
University of Science and Technology MISIS,\par
6~Leninskiy prospekt, Moscow, 119991, Russia;}
\affil[*]{vadim.petrukhanov@gmail.com, \href{http://www.mathnet.ru/eng/person176798}{mathnet.ru/eng/person176798}}
\affil[**]{apechen@gmail.com, \href{http://www.mathnet.ru/eng/person17991}{mathnet.ru/eng/person17991}}

\maketitle

\begin{abstract}In this work, we consider an environment formed by incoherent photons as a resource for controlling open quantum systems via an incoherent control. We exploit a coherent control in the Hamiltonian and an incoherent control in the dissipator which induces the {\it time-dependent decoherence rates} $\gamma_k(t)$ (via time-dependent spectral density of incoherent photons) for generation of single-qubit gates for a two-level open quantum system which evolves according to the Gorini--Kossakowski--Sudarshan--Lindblad (GKSL) master equation with time-dependent coefficients determined by these coherent and incoherent controls. The control problem is formulated as minimization of the objective functional, which is the sum of Hilbert-Schmidt norms between four fixed basis states evolved under the GKSL master equation with controls and the same four states evolved under the ideal gate transformation. The exact expression for the gradient of the objective functional with respect to piecewise constant controls is obtained. Subsequent optimization is performed using a gradient type algorithm with an adaptive step size that leads to oscillating behaviour of the gradient norm vs iterations. Optimal trajectories in the Bloch ball for various initial states are computed. A relation of quantum gate generation with optimization on complex Stiefel manifolds is discussed. We develop methodology and apply it here for unitary gates as a testing example. The next step is to apply the method for generation of non-unitary processes and to multi-level quantum systems.

\end{abstract}

\section{Introduction}

Optimal control of quantum systems has various applications ranging from quantum computing to optimal molecule discrimination~\cite{KochEPJQuantumTechnol2022,TannorBook2007, BrifChakrabartiRabitz2010,Moore2011,DAlessandroBook2021}. Often controlled quantum systems are interacting with the environment. Hence, they have to be considered as open quantum systems. In~some situations, the environment prevents from controlling the system. In~other cases, it can be used as a useful resource. An example is via {\it incoherent control} proposed in~\cite{Pechen_Rabitz_2006,PechenOSA2012}, where time dependent decoherence rates induced by non-equilibrium spectral density of incoherent photons are generally used as control jointly with coherent control by lasers.  This incoherent control uses the engineered environment as a useful resource.  Such incoherent control was applied to general multilevel systems in the dynamical picture (i.e., using master equations for the description of the dynamics) ~\cite{PechenPRA2011}, as well as to specific problems for single-qubit and two-qubit models, e.g.,~in~\cite{MorzhinPechenIJTP2021,Lokutsievskiy_2021,Morzhin_Pechen_LJM2021,Petruhanov2022}, etc. Controllability of open quantum systems in the kinematic picture (using universally optimal Kraus maps for description of evolution of the system) was studied in~\cite{Wu_2007_5681}. The~first experimental realization of Kraus maps studied in~\cite{Wu_2007_5681} for an open single qubit was done in~\cite{Zhang_Saripalli_Leamer_Glasser_Bondar_2022}.

Engineered environments were used for manipulation of quantum systems in various contexts. Preparation of mixed states and generation of non-unitary evolutions are necessary for quantum computing with mixed states and non-unitary circuits~\cite{Aharonov_1998,Tarasov_2002}. Engineered environments were suggested for improving quantum state engineering~\cite{Cirac2009}, cooling of translational motion without reliance on internal degrees of freedom~\cite{Calarco2011}, preparing entangled states~\cite{Diehl_nature_2008,Weimer_nature_2010}, improving quantum computation~\cite{Cirac2009}, inducing multiparticle entanglement dynamics~\cite{Barreiro_nature_2010}, and making robust quantum memories~\cite{Pastawski_2011}. A~scheme for steady-state preparation of a harmonic oscillator in the first excited state using dissipative dynamics was proposed~\cite{Borkje2014}. Quantum harmonic oscillator state preparation by reservoir engineering was considered~\cite{Kienzler2015}. Optimal control for cooling a quantum harmonic oscillator by controlling its frequency was proposed~\cite{Salamon2012}. Reachable states for a qubit driven by coherent and incoherent controls were analytically described using geometric control theory~\cite{Lokutsievskiy_2021}. Controllability analysis of quantum systems immersed within an engineered environment was performed~\cite{Grigoriu_2012}. Coherent control and simplest switchable noise on a single qubit were shown to be capable of transferring between arbitrary $n$-qubit quantum states~\cite{Bergholm_2012}. Quantum state preparation by controlled dissipation in finite time was proposed~\cite{Baggio_2012}. Precision limits to quantum metrology of open dynamical systems through environment control were studied~\cite{Davidovich_2013}.

Various optimization methods are used to find controls for quantum systems including Krotov type approaches (\cite{Tannor1992, Morzhin_Pechen_2019,Goerz_2014}, Zhu--Rabitz~\cite{ZhuRabitz1998} method, GRadient Ascent Pulse Engineering (GRAPE)~\cite{Khaneja2005}, 
Hessian based optimization in the Broyden--Fletcher--Goldfarb--Shanno (BFGS) algorithm and combined approaches~\cite{DalgaardPRA2020, deFouquieres2011}, 
speed gradient method~\cite{PechenBorisenokFradkov2022}, 
gradient free Chopped RAndom Basis (CRAB) optimization~\cite{CanevaPRA2011}, genetic algorithms~\cite{JudsonRabitz1992, Pechen_Rabitz_2006}, dual annealing~\cite{Morzhin_Pechen_LJM2021}, machine learning~\cite{Dong_Chen_Tarn_Pechen_Rabitz_2008}, etc. Quantum speed limit for a controlled qubit moving inside a leaky cavity has been studied~\cite{photonics9110875}. Preparation of single qubit and two-qubit gates has been studied, e.g.,~in~\cite{Fonseca2003,Grace2007,Zhang2013,Goerz_2014,Malinovsky2014,Ghaeminezhad2018,Li2022,Hegde2022,Jandura2022}, etc.

In this work, the~incoherent control in the dissipator is implemented using the environment formed by incoherent photons and is used as a resource, together with the coherent control in the Hamiltonian, to~control a single qubit system, based on the approach in~\cite{Pechen_Rabitz_2006}. The~incoherent control induces \textbf{time-dependent decoherence rates} $\gamma_k(t)$ via time-dependent spectral density of the environment $n(t)$, in~addition to the coherent control. The following master equation for the system density matrix $\rho(t)$ evolution was proposed~\cite{Pechen_Rabitz_2006}:
\begin{equation}\label{eq:ME}
\frac{d\rho(t)}{dt} = -i[H_0 + H_c(t),\rho(t)] + \sum\limits_k \gamma_k(t) {\cal D}_k(\rho(t))
\end{equation}

Here, $H_0$ is a free system Hamiltonian, $H_c(t)$ is a Hamiltonian describing interaction of the system with coherent control, $k$ denotes all possible different pairs of energy levels in the system, and ${\cal D}_k$ is a Gorini--Kossakowski--Sudarshan--Lindblad (GKSL) dissipator (we set Planck constant  as $\hbar=1$). In~addition to general consideration, two physical classes of the environment were exploited in~\cite{Pechen_Rabitz_2006} --- incoherent photons and quantum gas, with~two explicit forms of ${\cal D}_k$ derived in the weak coupling limit (describing atom interacting with photons) and low density limit (describing collisional type decoherence), respectively. In~\cite{PechenPRA2011}, it was shown that for the master equation~(\ref{eq:ME}) with ${\cal D}_k$ derived in the weak coupling limit, a generic $N$-level quantum system becomes approximately completely controllable in the set of density~matrices.

As particular control example, we consider here generation of the single-qubit gate $X$ (Pauli matrix $X$) and Hadamard gate $H$ for a two-level open quantum system, evolving according to the Gorini--Kossakowski--Sudarshan--Lindblad master equation with time-dependent coefficients determined by these coherent and incoherent controls. The~control problem is formulated as the minimization of the objective functional, which is the sum of Hilbert--Schmidt distances between density matrices evolving under controls from some basis states for some time $T$ and target density matrices. The target density matrices are produced from the basis states using the ideal quantum gate. Note that using three specific mixed states is enough in the general case of a $N$-level quantum system~\cite{Goerz_2014}. We obtain an exact expression for the gradient of the objective functional with respect to piecewise constant controls and perform the optimization using a gradient type algorithm with an adaptive step size.  The~use of the adaptive step size leads to oscillating behavior of the norm of the gradient vs number of iterations, which is in contrast to Figure~2 in~\cite{PechenTannorIJC2012}. In that figure, the norm of the gradient of the objective first monotonically increases, and then, monotonically decreases with number of the iterations. We visualize the results by plotting in the Bloch ball evolution of the initial states under obtained controls which are in a good agreement with the ideal~case.

We utilize the gradient descent method for optimization. In the context of quantum control, it is known as GRAPE. The~gradient descent is a local method and does not necessarily converge to a global minimum. It can converge to a local minimum if~such a local minimum exists. Therefore, for objective functionals with traps, i.e.,~local but not global minima, using slower global stochastic methods might be preferable. The~analysis of traps for quantum systems was initiated in~\cite{Rabitz_Hsieh_Rosenthal_science_2004} and studied for various models in~\cite{PechenTannorIJC2012,Pechen_Tannor_2011,FouquieresSchirmer_2013,PechenTannorCJC2014,Larocca2018,Zhdanov2018,Russell2018,PechenIl'in2016,Volkov_Morzhin_Pechen2021,DalgaardPRA2022}, etc.

The analysis of this work is related to the approach developed in~\cite{PechenJPA2008,OzaJPA2009} to control of open quantum systems, where Kraus maps are described by points of a {\it complex Stiefel manifold} (strictly speaking, factor of the Stiefel manifold over some equivalence relation).  This picture is called a {\it kinematic picture} for description of the dynamics. In~these works, the~gradient and Hessian based optimization methods were developed for optimization of control objectives for open quantum systems. In~the present work, we consider the {\it dynamical picture}, where evolution of an open qubit is described by a master equation. The~relation between the two approaches is that this master equation induces some time evolved Kraus map, which corresponds to some trajectory on a suitable factor of  a Stiefel manifold. The gradient of the objective in the dynamic picture, which is used to find optimal controls, can be computed using chain rule for derivatives and gradient of the objective computed in the kinematic~picture. 

The structure of this paper is the following. In~Section~\ref{Sec2:ME}, the~master equation for a qubit driven by coherent and incoherent photons is provided. In~Section~\ref{Sec:IC}, incoherent control by photons is discussed in more details. In~Section~\ref{Sec3:Obj}, the~objective functional which describes the problem of generation of single-qubit quantum gates is formulated. In~Section~\ref{Sec4:GR}, the gradient of the objective is computed for piecewise constant controls. In~Section~\ref{Sec5:Num}, numerical optimization results are provided, including plots for optimal coherent and incoherent controls, and~for the evolution of the qubit basis states in the Bloch ball under the action of the optimal controls. In~Section~\ref{Sec:Discussion}, a connection of the obtained results with gradient optimization over Stiefel manifolds~\cite{PechenJPA2008,OzaJPA2009} is discussed. The conclusions in  Section~\ref{Sec6:Concl} summarizes the~results.

\section{Master Equation for a Qubit Driven by Coherent and Incoherent~Controls}\label{Sec2:ME} 

We consider a master equation for an open two-level quantum system (qubit) driven by coherent and incoherent controls with time-dependent decoherence rate determined by $\gamma(t) = \gamma [n(t)+1/2]$, where $n(t)$ is the spectral density of incoherent photons surrounding the qubit at the time~$t$:
\begin{equation}
    \frac{d\rho}{dt}  = - i [H_0 + V u(t), \rho] + \gamma \mathcal{L}_{n(t)}(\rho),\quad \rho(0) = \rho_0.
    \label{system_qubit}
\end{equation}
Here, $H_0 = \omega \begin{pmatrix}
0 & 0 \\
0 & 1
\end{pmatrix}$ is the free Hamiltonian, 
$V = \mu\sigma_x
= \mu\begin{pmatrix}
0 & 1 \\
1 & 0
\end{pmatrix}$ is the interaction Hamiltonian, $\omega$ is the transition frequency, $\mu>0$ is the dipole moment, $u(t)$ is coherent control (real-valued function), and $\gamma>0$ is the decoherence rate coefficient. The~dissipative superoperator is
\begin{multline*} 
\mathcal{L}_{n(t)}(\rho_t) = n(t) \left(\sigma^+\rho_t\sigma^- + \sigma^-\rho_t\sigma^+ - \dfrac{1}{2}\{\sigma^-\sigma^+ + \sigma^+\sigma^- , \rho_t\}\right) \\ + \left(\sigma^+\rho_t\sigma^- -  \dfrac{1}{2}\{\sigma^-\sigma^+, \rho_t\}\right),
\end{multline*}
where $n(t) \geq 0$ is incoherent control (non-negative real-valued function), matrices $\sigma^\pm = \dfrac{1}{2} (\sigma_x \pm i\sigma_y)$, $\sigma^+ = \begin{pmatrix}
0 & 1 \\
0 & 0
\end{pmatrix}$, $\sigma^-= \begin{pmatrix}
0 & 0 \\
1 & 0
\end{pmatrix}$, and~$\sigma_x$, $\sigma_y$, $\sigma_z$ are the Pauli~matrices.

We use convenient parameterization of the density matrix~$\rho$ by Bloch vector~$\mathbf{r} = (r_x, r_y, r_z) \in\mathbb R^3$:
\begin{equation}
    \rho = \dfrac{1}{2} \Bigg(\mathbb{I} + \sum_{\alpha \in \{x, y, z\}}r_\alpha\sigma_\alpha\Bigg), \quad r_\alpha = \mathrm{Tr}\rho \sigma_\alpha,\quad \alpha \in \{x, y, z\},  
    \label{bloch variables}
\end{equation}
where $\mathbb{I}$ is the $2\times2$ identity matrix. In~this representation, we obtain the following equation for the dynamics of the Bloch vector $\mathbf{r}(t)$:
\begin{equation}
    \frac{d\mathbf{r}}{dt} = A(t) \mathbf{r} + \mathbf{b} =  \left(B + B^u u(t) + B^n n(t) \right) \mathbf{r} + \mathbf{b},\quad \mathbf{r}(0) = \mathbf{r}_0,
    \label{bloch equation}
\end{equation}
where $r_{0\alpha} = \mathrm{Tr}\rho_0 \sigma_\alpha$, $\alpha \in \{x, y, z\},$ and
$$ 
B = \begin{pmatrix}
- \dfrac{\gamma}{2} & \omega & 0 \\
- \omega & - \dfrac{\gamma}{2} & 0 \\
0 & 0 & - \gamma \\
\end{pmatrix}, B^u = 2\mu
\begin{pmatrix}
0 & 0 & 0 \\
0 & 0 & -1 \\
0 & 1 & 0 \\
\end{pmatrix},  B^n = -\gamma
\begin{pmatrix}
1 & 0 & 0 \\
0 & 1 & 0 \\
0 & 0 & 2 \\
\end{pmatrix}, \mathbf{b} = 
\begin{pmatrix}
0 \\
0 \\
\gamma \\
\end{pmatrix}.$$

\section{Incoherent~Control}\label{Sec:IC}

\begin{figure}[ht!]
\centering
\includegraphics[width = 0.6\linewidth]{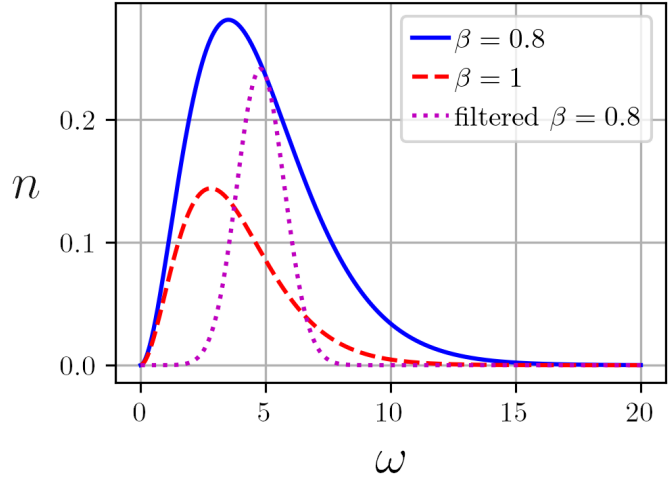}
\caption{Planck density of black-body radiation for $\beta = 0.8, 1$ and Gaussian filtering for $\beta = 0.8$
centered at $\omega=5$ with a variance $\sigma^2=1$.}
    \label{Fig1}
\end{figure}

Incoherent control as proposed in~\cite{Pechen_Rabitz_2006} is realized by using the environment as a resource to manipulate quantum systems. Physically incoherent control can be realized by using incoherent light and tailoring its distribution in frequency. In~this case, control is density of photons $n_\omega(t)$ with frequency $\omega=|{\bf k}|c$ (${\bf k}$ is photon momentum and $c$ is the speed of light) and time $t$. In~general, photons have polarization $\alpha$, and their proper distribution should include dependence on polarization, if~any, so that incoherent control becomes a function of $\omega$, $\alpha$, and~$t$. However, we neglect this dependence in the present work. For~thermal light or black-body radiation, distribution of its photons has the famous Planck form
\[
n(\omega)= \frac{1}{\pi^2}\dfrac{\omega ^3}{\exp{(\beta \omega)} - 1},
\]
where $\beta$ is the inverse temperature (in the Planck system of units, where reduced Planck constant $\hbar$, speed of light $c$, and~Boltzmann constant $k_\textrm{B}$ equal one). However, it can generally be non-thermal with other shape $n_\omega(t)$, which can evolve in time. One can obtain non-thermal shapes, e.g.,~by filtering of black-body radiation, using light-emission diodes, etc. A~plot with Planck distribution and some of its filterings are shown in Figure~\ref{Fig1}.

Since physical meaning of incoherent control is density of particles of the environment, mathematically, it is a non-negative quantity, $n_\omega(t)\ge 0$. In~general, one can require that 
\[
\int\limits_0^\infty n_\omega(t)d\omega<\infty\qquad \forall t \geq 0,
\]
meaning that total density of photons at any time moment is finite. Decoherence rates for a quantum system, e.g.,~an atom, interacting with incoherent light, are determined by strength of interaction between the atom and light, and~by this density of photons. Such incoherent control naturally makes decoherence rates time dependent. The magnitude of the decoherence rates affects the speed of decay of off-diagonal elements of the system density matrix and determines the values of the diagonal elements towards which the diagonal part of the system density matrix evolves. This property can be used, for~example, for~approximate generation of various density matrices~\cite{PechenPRA2011}. 

In our case of a single qubit, the~qubit only has one transition frequency $\omega_0$. In the first approximation, the spectral density of incoherent light at only this frequency, $n(t)=n_{\omega_0}(t)$, affects the system. Thus, in this case, without loss of generality, one can use only thermal radiation with variable temperature. For~multilevel systems, this is not the case and thermal distribution with variable temperature does not provide the full generality for incoherent~control at all system's transition frequencies.

\section{Objective Functional for Single-Qubit Gate~Generation}\label{Sec3:Obj} 

Let us introduce the evolution superoperator $\Phi(t, u, n)$, also called as dynamical map, which is a completely positive and trace preserving superoperator. The~evolution superoperator $\Phi(t, u, n)$ acts on the initial density matrix $\rho_0$ and maps it to the solution of the system~(\ref{system_qubit}) evolving under controls $u$ and $n$ for $t \geq 0$:
\begin{equation}
    \rho(t, u, n) = \Phi(t, u, n) \rho_0.
    \label{evolution_operator}
\end{equation}
To find a control that steers the initial state $\rho_0$ to the target state $\rho_\textrm{target}$ as close as possible, one can use the Hilbert--Schmidt distance
\begin{equation}
    J(\rho, \rho_\textrm{target}) = \|\rho - \rho_\textrm{target}\|^2 = \mathrm{Tr} (\rho - \rho_\textrm{target})^2,
    \label{Hilbert-Schmidt distance}
\end{equation}
and solve the optimal control problem of finding controls $u$ and $n$ which minimize the objective functional
\begin{equation}
    F(u, n) = J\left(\Phi(T, u, n) \rho_0, \,\rho_\textrm{target}\right) = \|\Phi(T, u, n) \rho_0 - \,\rho_\textrm{target} \|^2\to \inf.
    \label{problem_minimize_distance}
\end{equation}

It should be noted that there might be many different controls that steer the system to the given target~state.

Generation of a unitary quantum gate $U$ means that we are looking for controls $u$ and $n$ which produce the evolution operator $\Phi(T, u, n)$ that is close enough to the "target" dynamic map $U\cdot U^\dagger$. Hence, we need to use a functional which will describe similarity of two dynamic maps. Let $\rho^{(j)}$, $j = 1,\dotsc,N$ be the basis density matrices in the space of $N\times N$ density matrices. Two linear maps coincide if they act identically on the basis. Therefore, the following functional can be used to compare two dynamic maps $\mathcal{W}$ and $\mathcal{W\,'}$:
\begin{equation}
    D_{\rho_0^{(1)}, \dotsc, \rho_0^{(N)}}(\mathcal{W},\mathcal{W\,'}) = \frac{1}{N}\sum_{j = 1}^N J\left(\mathcal{W} \rho_0^{(j)}, \,\mathcal{W\,'}\rho_0^{(j)}\right).
    \label{maps_functional}
\end{equation}

Hence, for gate generation, we consider the objective functional
\begin{multline}
    F_{U,N}\left(u, n; \rho_0^{(1)}, \dotsc, \rho_0^{(N)}\right) = D_{\rho_0^{(1)}, \dotsc, \rho_0^{(N)}}(\Phi(T, u, n),\, U\cdot U^\dagger) = \\
    \frac{1}{N}\sum_{j = 1}^N J\left(\Phi(T, u, n) \rho_0^{(j)}, \,U\rho_0^{(j)}U^\dagger\right) = \frac{1}{N}\sum_{j = 1}^N \left\| \rho(T, u, n) - U\rho_0^{(j)}U^\dagger \right\|^2\to \inf.
    \label{problem_gate_generation}
\end{multline}

Now, we formulate this problem using the Bloch parameterization. Make the correspondence between density matrices and vectors in the Bloch ball as
\begin{align}
    \rho(t, u, n) &\mapsto\mathbf{r}^{(j)}(t, u, n),\\
    U\rho_0^{(j)}U^\dagger &\mapsto\mathbf{r}^{(j)}_\textrm{target}.
\end{align}

Then, the problem~(\ref{problem_gate_generation}) is formulated as
\begin{equation}
    F_{U,N}\left(u, n; \rho_0^{(1)}, \dotsc, \rho_0^{(N)}\right) = 
    \frac{1}{2 N}\sum_{j = 1}^N \left|\mathbf{r}^{(j)}(T, u, n) -  \mathbf{r}^{(j)}_\textrm{target} \right|^2 \to \inf. 
    \label{problem_gate_generation_Bloch}
\end{equation}

It can be seen that values of the objective functional (\ref{problem_gate_generation_Bloch}) are bounded, both below and above, as
\begin{equation*}
    0 \leq F_{U,N}\left(u, n, \rho_0^{(1)}, \dotsc, \rho_0^{(N)}\right) \leq 2.
\end{equation*}

Finding exact maximum and minimum values for a fixed time $T$, basis $\rho_0^{(j)}$ and gate $U$ is not a trivial~problem.

We use $N = 4$ initial pure states $\rho_0^{(j)}$, $j = 1,2,3,4$, which form a basis in the linear space of $2\times2$ Hermitian matrices, to~define action of the unitary gate $U$:
\begin{equation}
\begin{split}
    \rho_0^{(1)} &= \ket{0}\bra{0} = 
    \begin{pmatrix}
    1 & 0 \\
    0 & 0
    \end{pmatrix}, \\ 
    \rho_0^{(2)} &= \ket{1}\bra{1} = 
    \begin{pmatrix}
    0 & 0 \\
    0 & 1
    \end{pmatrix},\\ 
    \rho_0^{(3)} &= \ket{+}\bra{+} = \frac{1}{2}
    \begin{pmatrix}
    1 & 1 \\
    1 & 1
    \end{pmatrix},\\ 
    \rho_0^{(4)} &= \ket{i}\bra{i} = \frac{1}{2}
    \begin{pmatrix}
    1 & -i \\
    i & 1
    \end{pmatrix}.
\end{split}
\label{basis_states_system}
\end{equation}
Here, $\rho_0^{(j)}$ are pure state density matrices formed by eigenvectors of Pauli matrices $\sigma_z$ (vectors $\ket{0}$ and $\ket{1}$), $\sigma_x$ (vector $\ket{+} = (\ket{0} + \ket{1})/\sqrt{2}$), and~$\sigma_y$ (vector $\ket{i} = (\ket{0} + i\ket{1})/\sqrt{2}$).  Note that, as was mentioned in the Introduction, three special mixed states are enough in the general case of a quantum system of an arbitrary dimension~\cite{Goerz_2014}. 

Any single-qubit initial density matrix $\rho_0$ can be uniquely decomposed in the basis~(\ref{basis_states_system}):
\begin{multline}
    \rho_0 = 
    \begin{pmatrix}
        x_1 & x_3 + i~x_4 \\
        x_3 - i x_4 & x_2
    \end{pmatrix} 
    = (x_1 + x_4 - x_3) \rho_0^{(1)} + \\ + (x_2 + x_4 - x_3) \rho_0^{(2)} + 2 x_3\rho_0 ^{(3)} - 2 x_4 \rho_0^{(4)}.
    \label{initial_state_decompozition}
\end{multline}

Therefore, this number of states is sufficient to define any linear operator: particularly, any unitary transformation $\rho \mapsto U \rho U^\dagger$.
The corresponding Bloch vectors~$\mathbf{r}_0^{(j)}$ are
\begin{equation}
\begin{split}
    \mathbf{r}_0^{(1)} = (0, 0, 1),\quad 
    \mathbf{r}_0^{(2)} = (0, 0, -1),\quad
    \mathbf{r}_0^{(3)} = (1, 0, 0),\quad
    \mathbf{r}_0^{(4)} = (0, 1, 0).
\end{split}
\label{basis_states_system_Bloch}
\end{equation}

We also consider the objective functional $F_{U,2}$ with only two states $\rho_0^{(0)} = \ket{0}\bra{0}$ and $\rho_0^{(1)} = \ket{1}\bra{1}$ to show that this system is insufficient for generating quantum gates. Indeed, the~quantum gate $\sqrt{Y} \sim \exp(-i\pi/4Y)$ (rotation along y-axis by $\pi/2$ angle in the Bloch ball) acts exactly as the Hadamard gate $H$ on the states $\rho_0^{(0)}$ and $\rho_0^{(1)}$ :
\begin{align}
    \sqrt{Y}\ket{0}\bra{0}\sqrt{Y}^\dagger &= H\ket{0}\bra{0} H = \ket{+}\bra{+},\\
    \sqrt{Y}\ket{1}\bra{1}\sqrt{Y}^\dagger &= H\ket{1}\bra{1} H = \ket{-}\bra{-}.
    \label{two_insufficient}
\end{align}
However, these two gates are different.

\section{Gradient of the Objective~Functional}\label{Sec4:GR} 

The main features of quantum control landscape are local (if they exist) and global extrema of the objective functional. Since extrema are critical points where gradient of the objective is zero, an~important  step for the analysis of quantum control landscapes is the derivation of the expression for gradient of the objective. Below, we derive the gradient for piecewise constant control. Incoherent control is constrained: $n(t) \geq 0$; therefore, we make a variable change:
\begin{equation}
    n(t) = w(t)^2,
\end{equation}
so that control becomes a pair of $u$ and $w$.
The controls, $u(t)$ and $w(t)$, as piecewise constant functions, are defined by $M$-dimensional real vectors $u_k$ and $w_k$:
\begin{align}
    u(t) &= \displaystyle \sum_{k=1}^M u_k \chi_{[t_{k-1}, t_k)}(t), \qquad u_k\in\mathbb R\label{PConst qubit control: u}\\
    w(t) &= \displaystyle \sum_{k=1}^M w_k \chi_{[t_{k-1}, t_k)}(t), \qquad w_k\in\mathbb R\label{PConst qubit control: w}
\end{align}
where $0 < t_0 < t_1 < \dots < t_M = T,$ and $\chi_{[t_{k-1}, t_k)}$ is the characteristic function of the half-open interval $[t_{k-1}, t_k)$. For the sake of brevity, we denote this pair as $v = \left(v^1, v^2\right) = (u, w)$ and its components as $v_k = \left(v^1_k, v^2_k\right) = \left(u_k, w_k\right)$, $k = 1, \dotsc, M$.

The matrix $A(t)$ in the right-hand side (r.h.s.) of the Equation~(\ref{bloch equation}) also takes only $M$ different values $A_k$:
$$A_k = B + B^u u_k + B^n w_k^2,\quad k = 1, \dots, M.$$

The solution of the equation with piecewise constant r.h.s can be expressed using the following recurrent formula:
\begin{multline}
    \mathbf{r}_k \equiv \mathbf{r}(t_{k}) = e^{A_k \Delta t_k} \mathbf{r}_{k - 1} + \mathbf{g}_k = e^{A_k \Delta t_k} \dotsm e^{A_1 \Delta t_1} \mathbf{r}_0 \\+ e^{A_k \Delta t_k} \dotsm e^{A_2 \Delta t_2} \mathbf{g}_1  +\dotsb + e^{A_k \Delta t_k} \mathbf{g}_{k-1} + \mathbf{g}_{k},
    \label{Bloch k_th state}
\end{multline}
where
\begin{equation}
    \mathbf{g}_k = (e^{A_k \Delta t_k} - \mathbb{I}) A_k^{-1} \mathbf{b}, \quad \Delta t_k  = t_k - t_{k - 1}, \quad k = 1, \dots, M.
    \label{g definition}
\end{equation}

Differentiating the final state $\mathbf{r}_M = \mathbf{r}(T)$~(\ref{Bloch k_th state}) with respect to the
$k$th component of piecewise constant controls~$u$~(\ref{PConst qubit control: u}) and~$w$~(\ref{PConst qubit control: w}) gives
\begin{multline} 
    \pd{\mathbf{r}(T)}{v^m_k}  = e^{A_N \Delta t_N} \dots e^{A_{k + 1} \Delta t_{k + 1}}\bigg[ \pd{}{v_k^m}\left(e^{A_k \Delta t_k}\right)\mathbf{r}_{k-1} + \pd {\mathbf{g}_k}{(u_k, w_k)}\bigg],\\
    k = 1, \dots, M,\quad m = 1,2.
    \label{final state gradient u and w}
\end{multline}

The derivatives are given by
\begin{align}
    &\pd {\mathbf{g}_k}{v^1_k} = \pd {\mathbf{g}_k}{u_k} = \left(\pd{}{u_k}e^{A_k \Delta t_k} - (e^{A_k \Delta t_k} - \I)A_k^{-1} B^u\right) A_k^{-1} \mathbf{b},\label{g gradient u}\\ 
    &\pd{}{v^1_k}e^{A_k \Delta t_k} = \pd{}{u_k}e^{A_k \Delta t_k} = \int \limits _0^{\Delta t_k} e^{A_k t }\, B^u\, e^{A_k (\Delta t_k-t)} {\mathrm{d}} t,\label{F gradient u}\\
    &\pd {\mathbf{g}_k}{v^2_k} = \pd {\mathbf{g}_k}{w_k} = \left(\pd{}{w_k}e^{A_k \Delta t_k} - 2 w_k (e^{A_k \Delta t_k} - \I)A_k^{-1} B^n\right) A_k^{-1} \mathbf{b},\label{g gradient w}\\ 
    &\pd{}{v^2_k}e^{A_k \Delta t_k} = \pd{}{w_k}e^{A_k \Delta t_k} = 2 w_k \int \limits _0^{\Delta t_k} e^{A_k t }\, B^n\, e^{A_k (\Delta t_k-t)} {\mathrm{d}} t,\quad k = 1, \dotsc, M.\label{F gradient w}
\end{align} 

The expressions~(\ref{F gradient u}) and~(\ref{F gradient w}) were found using the integral formula~\cite{Wilcox}:
\begin{equation}
    \der{}{x} e^{A(x)} = \int \limits _0^1 e^{s A(x)} \der{A(x)}{x} e^{(1-s)A(x)} {\mathrm{d}} s.
    \label{special_formula}
\end{equation}

Finally, the~derivatives of the functional~(\ref{problem_gate_generation_Bloch}) with respect to the components $u_k$ and $w_k$ of the controls $u$ and $w$, respectively, are
\begin{equation}
    \pd{F_{U,N}}{v^m_k} = 
    \frac{1}{N}\sum_{j = 1}^N \left(\mathbf{r}^{(j)}(T) -  \mathbf{r}^{(j)}_\textrm{target}\right) \cdot \pd{\mathbf{r}^{(j)}(T)}{v^m_k},\quad k = 1, \dotsc, M,\quad m = 1,2.
    \label{gradient_objective_functional}
\end{equation}

\section{\texorpdfstring{Numerical Optimization for Generation of $X$ and $H$ Gates}
{Numerical Optimization for Generation of X and H Gates}}\label{Sec5:Num} 

In this section, we use the above derived gradient~(\ref{gradient_objective_functional}) of the objective function to implement a gradient optimization method for generating single-qubit quantum gates. We consider two single-qubit gates: the $X$ gate, which is $\sigma_x$ Pauli matrix, and~the Hadamard gate $H$, i.e.,~$U \in \{X, H\}$:
\begin{equation}
    X = \sigma_x = 
\begin{pmatrix}
    0 & 1 \\
    1 & 0
\end{pmatrix},\quad
H = \frac{1}{\sqrt{2}}\begin{pmatrix}
    1 & 1 \\
    1 & -1
\end{pmatrix}.
\label{two_gates}
\end{equation}

To find controls $u$~(\ref{PConst qubit control: u}) and $w$~(\ref{PConst qubit control: w}) that minimize the objective functional~(\ref{problem_gate_generation_Bloch}), we utilize the gradient descent method known in the context of quantum control as GRAPE. This is an iterative algorithm, where each update of the control $v = (u, w)$ describes moving against the gradient of the objective functional:
\begin{equation}
    v^{(l + 1)} = v^{(l)} - h^{(l)} \mathrm{grad}_{v} G_{U}\big(v^{(l)}\big),\quad l = 0,1,\dotsc;
    \label{gradient_descent}
\end{equation}

Here, for brevity, we denote
\begin{equation}
    G_{U}\left(v\right) = G_{U}\left((u, w)\right) = F_{U,4}\left(u, w^2; \rho_0^{(1)}, \rho_0^{(2)},\rho_0^{(3)}, \rho_0^{(4)}\right),
    \label{functional_on_v}
\end{equation}
where the basis matrices $\left\{\rho^{(j)}\right\}_{j = 1}^{4}$ are fixed and chosen as~(\ref{basis_states_system}), their Bloch representations are given by~(\ref{basis_states_system_Bloch}); $\mathrm{grad}_{v}$ denotes a $2M$-dimensional vector operator, each of the components acts according to formula~(\ref{final state gradient u and w}); $v^{(l)} = \left(u^{(l)}, w^{(l)}\right)$ is the control on the $l$th iteration; and $h^{(l)}$ is the size of the $l$th~step.

The stopping criterion is the condition for the fidelity to be small enough, such that fidelity value on the $l$th step is less than some precision parameter $\varepsilon_1  > 0$:
\begin{equation}
    G_{U}^{(l)} = G_{U}\big(v^{(l)}\big) < \varepsilon. \label{fidelity_stop_criterion}
\end{equation}

This stopping criterion is satisfied when we find an optimal control $v = (u,w)$ that steers the initial basis states $\rho_0^{(j)}$ to the final states $\Phi\left(t, u, w^2\right)\rho_0^{(j)}$ that are different from the action of the unitary gate $U$ on them, $U\rho_0^{(j)}U^\dagger$, less than $\varepsilon_1$ in the sense of mean square of the Hilbert--Schmidt distances~(\ref{problem_gate_generation}). Since each control $v = (u, w)$ defines some dynamical map $\Phi\left(t, u, w^2\right)$ on the space of density matrices, we attempt to construct a dynamical map which is as close as possible to the desired map $U \cdot U^\dagger$ produced by the target quantum gate $U$.

Performance of gradient descent~(\ref{gradient_descent}) largely depends on the choice of a step size sequence~$h^{(l)}$. Often with a predefined sequence~$h^{(l)}$, gradient descent method may produce a fidelity value sequence~$G_{U}^{(l)}$ which is not monotonically decreasing. We propose the adaptive choice of step size and modification of the control update rule~(\ref{gradient_descent}). In~each $(l+1)$th iteration,  we compute the potential next updated value of fidelity $G_U\left(v^{(l)} - h^{(l)}\mathrm{grad}_v G\big(v^{(l)}\big)\right)$ and compare it with the last value on the $l$th iteration $G_{U}^{(l)}$; if~it is lower, we update the control; otherwise, the control and the fidelity value remain the same:
\begin{equation}
    v^{(l+1)} = 
    \begin{cases}
        v^{(l)} - h^{(l)}\mathrm{grad}_v G\big(v^{(l)}\big) & \quad \text{if } G_U\left(v^{(l)} - h^{(l)}\mathrm{grad}_v G\big(v^{(l)}\big)\right) <\\
        & \qquad\qquad G_{U}\big(v^{(l)}\big);\\
        v^{(l)}& \quad \text{o/w.} 
    \end{cases}
    \label{gradient_descent_modified}
\end{equation}

Increase of the fidelity means that the step size is too large; therefore, we decrease the next step in the second case. We also increase the step size in the first case to boost up the gradient descent:
\begin{equation}
    h^{(l+1)} = 
    \begin{cases}
        c h^{(l)} & \quad \text{if } G_U\left(v^{(l)} - h^{(l)}\mathrm{grad}_v G\big(v^{(l)}\big)\right) <\\
        & \qquad\qquad G_{U}\big(v^{(l)}\big);\\
        d h^{(l)} & \quad \text{o/w;} 
    \end{cases}
    \label{gradient_descent_steps_modified}
\end{equation}
where $c \geq 1$, $0 < d < 1.$ 

Thus, we produce a monotonically (not strictly) decreasing sequence of fidelity values~$G_U^{(l)}$. This scheme allows us to adapt step size to the local change of the functional landscape. This approach may significantly reduce the number of iterations needed to reach the desired accuracy of optimization. The~adaptive scheme~(\ref{gradient_descent_steps_modified}) depends on the three parameters, which were chosen as: $h^{(0)} = 1$, $c = 1.1$, $d = 0.5$.

The important question is whether the adaptive scheme~(\ref{gradient_descent_modified})~and~(\ref{gradient_descent_steps_modified}) may stuck in a non-critical point $x$ (i.e., in~a point with gradient $\nabla f(x) \neq 0$), i.e.,~may constantly reduce size of the step without decrease of the objective functional. If~gradient is calculated with absolute precision, there would exist a step size $h$ which is sufficient to decrease the value of the objective functional. This intuitive statement can be proved by the following simple reasoning. Consider a differentiable functional $F: D \subset \R^n \to \R$. Then, in a vicinity of an internal point $x \in D$:
\begin{equation*}
    F(x - h \nabla F(x)) - F(x) = -h |\nabla F(x)|^2 + o(h),\quad h \to +0.
\end{equation*}

Due to the definition of $o(h)$, for~any $\epsilon > 0$, there exists $h > 0$ such that $o(h) < \epsilon h$. If~$x$ is not a critical point (so that $\nabla F(x)\ne 0$), we can take $\epsilon = |\nabla F(x)|^2$ and $F(x - h \nabla F(x)) < F(x)$. Thus, the answer to the posed question is ''no'' --- after a finite number of iterations~(\ref{gradient_descent_steps_modified}), the step size will be small enough to decrease the objective functional value. However, practice of numerical experiments shows that the adaptive scheme sometimes does not provide the objective functional value decrease over quite large number of iterations. In~order to deal with such cases, we utilize one more stop criterion. The~second criterion is the following: if $G_U^{(l)} = G_U^{(r)},  r = l - 1, l - 2, \dotsc, l - L_\textrm{stuck}, l \geq L_\textrm{stuck}$, the iterations also are stopped. We set $L_\textrm{stuck} = 20$ as the maximum number of "stationary"~iterations.

Numerical computations were performed by writing a Python program using the Numpy library for fast matrix operations and the SciPy library function \texttt{scipy.linalg.expm} for matrix exponentials computation using the Pad\'{e} approximation. The~following values of the system parameters were used: the transition frequency $\omega = 1$, the~dipole moment $\mu = 0.1$, and~the decoherence rate coefficient $\gamma = 0.01$. We use regular partition of the time segment~$[0, T]$ with $T = 5$ into $M = 10$ segments, such that each segment has length $\Delta t_k = T/M = 0.5$. Integral formulae were approximated by the trapezoidal rule with the number of partitions $N_{\rm partition} = 20$.

Initial guesses for a coherent control $u^{(0)}$ and for an incoherent control $w^{(0)}$ were chosen as discretization of the sine function and the normal distribution function (similar to physical realization of coherent and incoherent controls): $u^{(0)}_k = \sin{(2\pi t_{k-1}/T)}$, $n^{(0)}_k = {w^{(0)}_k}^2 = \exp{\left(-4 (t_{k-1}/T - 1/2)^2\right)}$. The~precision parameter was $\varepsilon = 0.001$.

\begin{figure}[ht!]
    \centering
    \includegraphics[width = \linewidth]{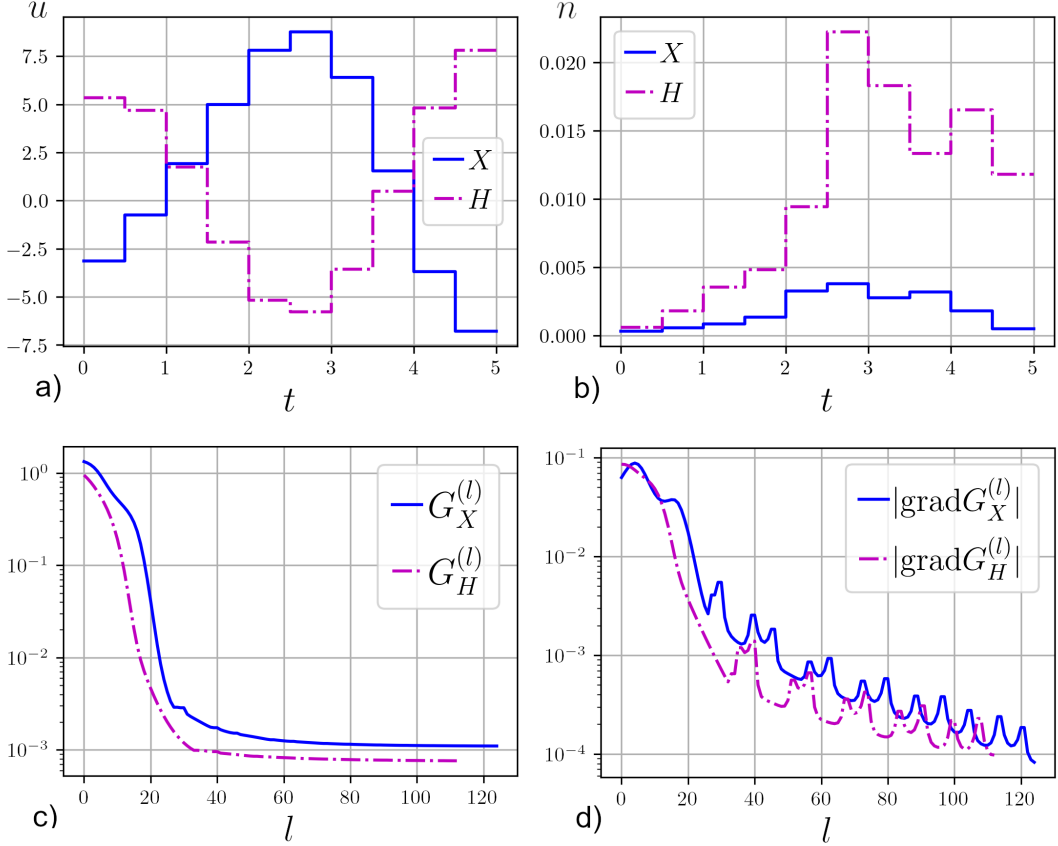}
    \caption{Optimized coherent control $u(t)$ (subplot \textbf{a}) and incoherent control $n(t)$ (subplot \textbf{b}) which generate     the quantum gate $X$ (blue line) and the Hadamard quantum gate $H$ (purple line); convergence of the fidelity --- the objective functional $G_U^{(l)},\;U \in \{X,H\}$ (\ref{fidelity_stop_criterion})--- (subplot \textbf{c}); norm of the gradient of the objective functional (subplot \textbf{d}) vs number $l$ of the optimization algorithm~(\ref{gradient_descent}) iterations.}
    \label{Fig2}
\end{figure}

Figure~\ref{Fig2} illustrates optimized coherent (subplot a) and incoherent controls (subplot b) for generating the gate $X$ (blue line) and the Hadamard gate $H$ (purple line), convergence of the fidelity --- the objective functional $G_U^{(l)},\;U \in \{X,H\}$, (\ref{fidelity_stop_criterion}) (subplot c)---and norm of the gradient of the objective functional (subplot d) vs number $l$ of the optimization algorithm~(\ref{gradient_descent}) iterations under the optimized coherent and incoherent controls. As~shown on subplot (a),  for~generation of $H$ gate, the fidelity goes to a small value of $10^{-3}$ in about $30$ iterations. For this gate, the convergence is fast. For~generation of $X$ gate, the~convergence to such a low value of the fidelity is much slower; even after $120$ iterations, the fidelity is still slightly above this value. Thus, high-precision generation of $X$ gate is more difficult with the present gradient approach. For~Hadamard gate, which is found to be easily generated with this approach, higher (compared to the case of $X$ gate) values of incoherent control were obtained, as~shown on subplot (b). The~difference of scales is large, since our target in this analysis is a unitary gate, and~as confirmed by the plots, it is natural to expect that incoherent control, while non-trivial, will have much smaller values than coherent. These gates are good testing examples and the results are in the agreement with expectation. The~next step will be to apply the method for generation of target non-unitary processes, where incoherent control component should be more significant, as~was found before for the problem of mixed state generation~\cite{Morzhin_Pechen_LJM2021}. 
Coherent controls, which were found here as the results of optimization, for~$X$ and $H$ gates are to some degree symmetric around a constant value $u\approx 2$.

\begin{figure}[ht!]
    \centering
    \includegraphics[width = \linewidth]{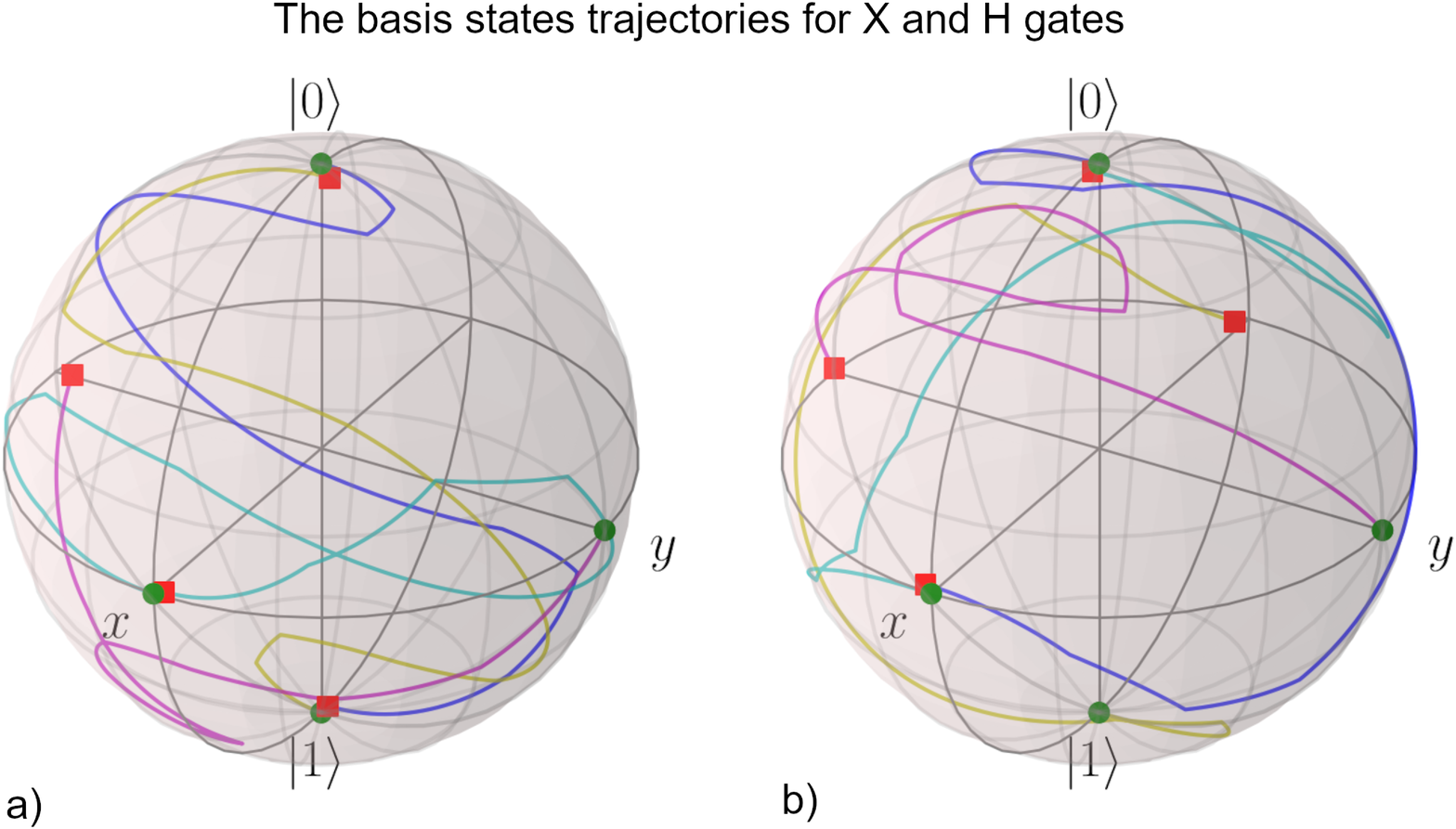}
    \caption{Optimal trajectories of the basis states~(\ref{basis_states_system_Bloch}) in the Bloch ball for generation of the quantum gate $X$ (subplot \textbf{a}) and the Hadamard quantum gate $H$ (subplot \textbf{b}). Initial states are indicated by green dots, final states by red squares, and~target states are centers of the black circles. The~trajectories lie close to the Bloch sphere, in~the region with a distance of about $0.05$ to the~sphere.}
    \label{Fig3}
\end{figure}

Figure~\ref{Fig3} shows optimal trajectories of the basis states~(\ref{basis_states_system_Bloch}) in the Bloch ball for generation of the quantum gate $X$ (subplot a) and the Hadamard quantum gate $H$ (subplot b), where four initial states, corresponding to $\rho_0^{(j)}$ ($j=1,2,3,4$) are indicated by green dots, final states by red squares, and~target states are centers of the black circles. Optimal trajectories are very close to the Bloch sphere (the longest distance to the sphere is up to $0.05$). However, they are located in the interior of the Bloch ball due to unavoidable non-unitary evolution of the system and the presence of incoherent~control.

We note that on Figure~\ref{Fig3}, the norm of the gradient has oscillating behavior vs number of iterations. This is in contrast to the case of~\cite{PechenTannorIJC2012} (see Figure~2 in that paper), where the norm of the gradient of the objective first monotonically increases, and then, monotonically decreases with the number of iterations. This difference in the behavior is expected to be due to the adaptive step size used in the present~work.

\section{Discussion: Quantum Gate Generation as Optimization Over Complex Stiefel~Manifolds}\label{Sec:Discussion}

Our analysis in this work is directly related to the gradient optimization approach which was developed for open quantum systems using a proposed formulation of completely positive trace preserving dynamics of open quantum systems as points of {\it complex Stiefel manifold} (rigorously speaking, as~points in quotient space of complex Stiefel manifolds over some equivalence relation). The~corresponding theory of gradient optimization over complex Stiefel manifolds was developed for two-level~\cite{PechenJPA2008} and general $N$--level quantum systems~\cite{OzaJPA2009}.  

Recall that the dynamics of density matrix of an $N$-level open quantum system can be described as a completely positive trace preserving map (Kraus map), which has an operator sum representation
\[
\Phi(\rho)=\sum\limits_{i=1}^{N^2} K_i\rho K_i^\dagger
\]
where $K_i$ are $N\times N$ complex matrices such that
\begin{equation}\label{eq:KrausOSR}
\sum\limits_{i=1}^{N^2} K_i^\dagger K_i=\mathbb I_N
\end{equation}
and $\mathbb{I}_N$ is the identity $N\times N$ operator. Denote  $N^3 \times N$ matrix
\[
S=\left(\begin{array}{c}
    K_1 \\
    K_2 \\
    \dots \\
    K_{N^2}
\end{array}\right)
\]
The condition~(\ref{eq:KrausOSR}) is equivalent to 
\[
S^\dagger S=\mathbb I_{N}
\]

The set of matrices which satisfy this condition forms a complex Stiefel manifold $V_N(\mathbb C^{N^3})$, which is the set of all orthonormal $N$-frames in $\mathbb C^{N^3}$. As~is well known, a~given physical evolution (Kraus map) has a non-unique Kraus map operator-sum representation. Thus, different sets of $\{K_i\}$ may describe the same physical evolution and one should take the appropriate equivalence class ${V_N(\mathbb C^{N^3})}{/\!\!\sim}$. This equivalence was taken into account in~\cite{OzaJPA2009}. In~particular, the Hadamard gate corresponds to the $8\times 2$ matrix 
\[
S_{H}=\left(\begin{array}{c}
    H \\
    0_{2\times 2} \\
    0_{2\times 2} \\
    0_{2\times 2}
\end{array}\right)
\]
which is a point of the complex Stiefel manifold $V_2(\mathbb C^{8})$. 

Coherent and incoherent controls which act on our one-qubit system (and more generally, on~any quantum system) induce some time evolving Kraus map $\Phi_t$, which induces some trajectory on the Stiefel manifold. Thus, optimization of gate generation can be formulated as optimization over complex Stiefel manifold. The~corresponding theory, including gradient and Hessian-based optimization, in~the kinematic picture was developed in~\cite{PechenJPA2008,OzaJPA2009}. In~our example, optimal controls plotted on Figure~\ref{Fig2} induce some trajectory on the Stiefel manifold. Since the dimension of the manifold is $2N^4 - N^2 = 28$, the~corresponding evolution can not be drawn as one can draw for evolution of single-qubit states on the Bloch ball, but~the Stiefel matrices for any moment in time can be~computed.

\section{Conclusions}\label{Sec6:Concl} 
In this work, we have exploited the coherent control and the environment formed by incoherent photons as a resource for generation of single-qubit gates in a two-level open quantum system evolving according to the Gorini--Kossakowski--Sudarshan--Lindblad master
equation with time-dependent coefficients. The~coherent control enters in the Hamiltonian, and the~incoherent control enters in the dissipator and induces time-dependent decoherence rates of the system. The~exact expression for the gradient of the objective functional with respect to piecewise constant controls has been obtained, and the optimization has been performed using the gradient type algorithm. The~optimal coherent and incoherent controls and the basis states trajectories in the Bloch ball have been computed. The~numerical optimization shows that generation of the $H$ gate is fast (a value of $10^{-3}$ is obtained in about $30$ iterations), giving low values of fidelity. For the $X$ gate optimization, it takes a much higher number of iterations ($120$ iterations were not enough). Thus, high-precision generation of $X$ gate is more difficult with the present approach, that might be due to the limits of controllability of the system~\cite{Lokutsievskiy_2021}. For~the Hadamard gate, higher values of incoherent control were obtained, compared to the case of the $X$ gate. The~coherent controls, obtained as the results of optimization for $X$ and $H$ gates, are found to some degree to be symmetric around a constant value $u_*\approx 2$. In this work, we use the adaptive step size of the gradient search to analyze the landscape. This leads to oscillating behavior of the norm of the gradient vs number of iterations; this contrasts Figure~2 in~\cite{PechenTannorIJC2012}, where the norm of the gradient of the objective firstly monotonically increases, and then, monotonically decreases with the number of iterations. While we use the particular model for the dissipative superoperator, the~approach can be extended to other cases as well. Here, we develop the theoretical basis of the approach and apply it for target unitary gates as a testing example. In~this case, the~incoherent control magnitude is obviously small. The~next step is to apply the method for generation of target non-unitary processes, where incoherent control component should be more significant, as~was found before for the problem of mixed state generation~\cite{Morzhin_Pechen_LJM2021}.

\vspace{6pt} 

\section*{Acknowledgements}
The new results in Sections~\ref{Sec:IC}--\ref{Sec5:Num} were obtained in Steklov Mathematical institute and supported by Russian Science Foundation under grant No.~22-11-00330,
\url{https://rscf.ru/en/project/22-11-00330/}, and~in Section~\ref{Sec:Discussion} in the University of Science and Technology MISIS under the federal academic leadership program ‘‘Priority 2030’’ (MISIS Grant No.~K2-2022-025)

\end{document}